\documentclass[sigconf]{acmart}

\usepackage{graphicx}
\usepackage{float}
\usepackage{enumitem}

\AtBeginDocument{%
  }

\copyrightyear{2025}
\acmYear{2025}
\setcopyright{acmlicensed}\acmConference[UbiComp Companion '25]{Companion of the 2025 ACM International Joint Conference on Pervasive and Ubiquitous Computing}{October 12--16, 2025}{Espoo, Finland}
\acmBooktitle{Companion of the 2025 ACM International Joint Conference on Pervasive and Ubiquitous Computing (UbiComp Companion '25), October 12--16, 2025, Espoo, Finland}
\acmDOI{10.1145/3714394.3756236}
\acmISBN{979-8-4007-1477-1/2025/10}

\begin{document}

\title[How Many Times Do People Usually Experience Different Kinds of Stressors]{How Many Times Do People Usually Experience Different Kinds of Stressors Each Day?}

\author{Sameer Neupane}

\email{sameer.neupane@memphis.edu}
\orcid{0000-0002-5739-2362}
\affiliation{%
  \institution{University of Memphis}
  \streetaddress{}
  \city{Memphis}
  \state{Tennessee}
  \country{USA}
  % \postcode{}
}

\author{Mithun Saha}

\email{msaha1@memphis.edu}
\orcid{0000-0003-1510-4307}
\affiliation{%
  \institution{University of Memphis}
  \streetaddress{}
  \city{Memphis}
  \state{Tennessee}
  \country{USA}
  % \postcode{}
}

\author{David M. Almeida}

\email{dalmeida@psu.edu}
\orcid{0000-0002-5233-8148}
\affiliation{%
  \institution{The Pennsylvania State University}
  \streetaddress{}
  \city{University Park}
  \state{Pennsylvania}
  \country{USA}
  % \postcode{}
}
\author{Santosh Kumar}

\email{santosh.kumar@memphis.edu}
\orcid{0000-0002-9273-0291}
\affiliation{%
  \institution{University of Memphis}
  \streetaddress{}
  \city{Memphis}
  \state{Tennessee}
  \country{USA}
  % \postcode{}
}

\renewcommand{\shortauthors}{Sameer, et al.}

\begin{abstract}

Understanding how frequently people experience different kinds of daily stressors is crucial for interpreting stress exposure and informing mental health care. But it can't be directly estimated from current assessment methods, such as diaries, end-of-day interviews, and ecological momentary assessments (EMA), that use sparse sampling to limit participant burden, and a structured response format for uniformity. In this paper, we utilize stressor data collected in a 100-day field study with 68 participants that adopted wearable-triggered prompts and a freeform format to solicit stressors soon after they occurred, but limited its prompts to a small subset to keep the burden low. We develop asymptotic models to estimate the latent frequency of different kinds of real-life stressors that address sample sparsity and sampling bias. We find that people experience 5.39 stressors per day, on average. The top three are related to work (1.76/day), health (0.59/day), and transportation (0.55/day). These estimates offer a principled benchmark for interpreting individual stressor loads. They can also inform mental health care treatments and interventions by establishing population-level baselines.

% to contextualize patient-reported stressors. and help achieve the right balance between stress intervention and interruptions.  
\end{abstract}

\begin{CCSXML}
<ccs2012>
   <concept>
       <concept_id>10003120.10003138.10011767</concept_id>
       <concept_desc>Human-centered computing~Empirical studies in ubiquitous and mobile computing</concept_desc>
       <concept_significance>500</concept_significance>
       </concept>
   <concept>
       <concept_id>10003120.10003121.10003122.10011750</concept_id>
       <concept_desc>Human-centered computing~Field studies</concept_desc>
       <concept_significance>500</concept_significance>
       </concept>
 </ccs2012>
\end{CCSXML}

\ccsdesc[500]{Human-centered computing~Empirical studies in ubiquitous and mobile computing}
\ccsdesc[500]{Human-centered computing~Field studies}

\keywords{Wearables; Stress Detection; Stressors; Stress Interventions}

\maketitle

\section{Introduction}
Compared to major life events, daily stressors have distinct and potent associations with psychological, mental, and physical health outcomes~\cite{kanner1981comparison,delongis1982relationship}. The National Study of Daily Experiences (NSDE), conducted 42,000 end-of-day interviews using the Daily Inventory of Stressful Events (DISE) approach~\cite{almeida2002daily, almeida2011combining} over 20 years, to collect sources of stress from more than 3,500 adults. This study revealed that health outcomes are determined more by the nature of stress experienced than the frequency or severity of exposure~\cite{stawski2013associations}. This new finding has been supported by multiple studies across different health outcomes~\cite{piazza2013affective,charles2013wear,sin2015affective}.  

To incorporate daily stressors in mental healthcare, providers need to understand whether a patient’s stressor frequency deviates from normative baselines, so they can use it for diagnosis, triage, and treatment. Yet, there is currently a lack of empirical reference points to interpret real-life stressor loads for different types of stressors. This problem has remained open due to a lack of data that captures all kinds of real-life stressors as soon as they occur.

Prevalent protocols for recording real-life stressors differ in when and how frequently participants are prompted to recall and record the stressors they have recently experienced. Annual and monthly surveys detect major stressors but often miss transient events~\cite{stressAmerica,chandola2022mental}. End-of-day interviews offer finer resolution yet remain prone to recall bias~\cite{almeida2002daily}. Interval-based EMAs prompt participants multiple times daily, reducing recall bias and improving the detection of momentary stressors~\cite{zawadzki2019understanding}.
AI-triggered prompts extend this approach further by using physiological signals to detect potential stress events and prompting users in real-time to report if they are experiencing a stressor, thus getting even closer to the actual occurrence of the stressor~\cite{neupane2024momentary}. However, to keep the participant burden low, they prompt participants only for a subset of the detected events. Therefore, deriving \emph{the latent upper bound on the number of stressors a person experiences in a day, assuming unlimited prompts,} requires applying an appropriate model to a data set of real-life stressors that can account for its sparse and potentially biased sampling.

In this paper, we apply a new asymptotic model to the data from 68 participants in a 100-day field study~\cite{neupane2024momentary}, where participants wore smartwatches and responded to wearable-triggered EMA prompts to confirm stressful events and record stressors using freeform text. We develop models to account for sampling sparsity, sampling bias, and diminishing return behavior, before fitting an exponential model for asymptotic estimation of the latent frequency. 
%the relationship between response frequency and stressor reports referred to as \emph{prompt efficiency} using an asymptotic function to estimate the upper bound on daily stressor frequency from sparse observations in the real world 
We find that saturation occurs at 5.39 stressors per day across the study population. We then apply the same framework to the most prevalent stressor types and report their relative frequency.
% We found that the relationship between responses and reported stressors saturates at approximately 5.39 stressors per day across the study population. 
Our modeling reveals that some stressors that appear less in observed data may actually occur more in daily life, and vice versa, highlighting hidden patterns masked by biased sampling. By estimating the latent frequency of stressors, our work provides a normative foundation that can support mental health practitioners in identifying at-risk individuals, optimizing intervention delivery, and tailoring treatment strategies for both general and clinical populations.

\section{Related Works}

\subsection{EMA Protocols for Stress Assessment}

EMAs are commonly used to capture real-time stress experiences by reducing recall bias, improving ecological validity, and supporting JITAI delivery~\cite{stone2007historical,heron2010ecological}. Effective EMA design involves sampling frequency, question content, and prompt timing~\cite{janssens2018qualitative}. To balance data utility and user burden, studies use strategies such as minimum time gaps (e.g., 30 minutes~\cite{smets2018large}), varying prompt frequencies~\cite{mccarthy2015experimental}, and participant-initiated reporting~\cite{yao2022does}. Prompt schedules range from evenly spaced~\cite{scott2017pilot} to probabilistic approaches (e.g., Poisson processes~\cite{rathbun2013survival}).
{As stress is a dynamic phenomenon that can arise at any instant}, some studies send frequent prompts, e.g., 12 per day in~\cite{smets2018large}—but only for a few days (5 days in~\cite{smets2018large}) due to high participant burden. 
These studies still show low yield: only 2.9\% of responses indicated high stress and 11.4\% indicated moderate stress.  Longer-duration studies reduce the frequency but then have a lower chance of catching participants in a stressful moment. For example, in the 56-day study in~\cite{booth2022toward}, only one EMA was triggered, resulting in 0.844 responses per day, of which fewer than 5\% indicated high stress and fewer than 25\% indicated some stress. With 5 daily responses,~\cite{smets2018large} had a chance of catching participants in a moderate, high, or extremely high-stress state $0.715 =5*(0.029+0.114)$ times per day, whereas with 0.844 responses per day,~\cite{booth2022toward} had a chance of catching participants in a some, a lot, or a great deal of stress state $0.253 =0.844*(0.05+0.25)$ times per day. These works only captured the stress state and did not collect the stressor.

\subsection{Assessment of Stressors in the field}
When stressors are collected through end-of-day interviews, participants typically report only those events that remain accessible in memory. Interval-based EMAs, which prompt participants throughout the day, reduce this recall bias by asking about more recent experiences. However, some degree of autobiographical memory decay may still occur~\cite{bradburn1987answering}.  Prior studies reflect this trade-off between recall delay and stressor capture: daily interviews yielded an average of 0.51 stressors~\cite{koffer2016stressor}; end-of-day smartphone surveys increased that to 1.28 stressors~\cite{ram2014examining}; interval EMAs administered every 45 minutes captured 1.82 stressors~\cite{zawadzki2019understanding}; and AI-triggered EMAs averaging 5.2 prompts/day captured 1.62 stressors~\cite{neupane2024momentary}. Participants recalled experiencing stress about 2.67 times per day in weekly surveys~\cite{neupane2024momentary}. 
%However, this estimate is limited by recall bias, capped frequency reporting, and lack of breakdown across stressor types.  
Existing studies primarily report findings based on observed samples. In contrast, this work estimates the true underlying frequency of stressors using asymptotic modeling, accounting for sparsity and biases in the sampling protocol.

\section{Methods}

\subsection{Relationship between Expected Number of Stressors and  Response Frequency}\label{sec:problem}

Let \( P_d \) and \( P_r \) denote the number of prompts issued and responded to per day by participants, respectively. Each response may consist of a valid stressor reported. Let \( \mathcal{S} = \{ S_{1}, S_{2}, \dots, S_{n} \} \) denote the set of all the stressors reported. Each stressor \( S_i \in \mathcal{S} \) is assigned to exactly one predefined stressor category \( C_k \in C = \{C_1, C_2, \dots, C_{N}\} \). 

%As the number of times people experience a stressor is limited and the effect of each stressor takes time to dissipate from people's memory,

We use \emph{response efficiency} to denote the number of prompt responses that result in the reporting of a valid stressor. Stressful events in a typical day are usually limited to a finite number. Our goal is to estimate this \textit{latent frequency} of stressors per day. To achieve a convergence to this latent frequency as the number of responses \( P_r \) becomes unbounded, we need to model the \emph{diminishing return} behavior as $P_r$ increases. Such a model should account for large within- and between-person variability in how individuals respond to and report real-life stressors. The model should be extensible to estimate the latent frequency of all stressors, as well as that of specific kinds of stressors that are reported more widely.

\subsection{Dataset}
Most existing stressor-related datasets consist of stressors collected throughout the day via prompts issued at scheduled or random times, or those collected via end-of-day or event-contingent prompts. As some stressors may fade away from memory more rapidly than others, these datasets are unlikely to capture all kinds of stressors, making them unsuitable for our modeling. Further, asymptotic modeling requires estimating the number of stressors likely to be reported when the prompts are responded to at any time of the day. Therefore, new types of datasets are needed. 

The recently collected MOODS dataset~\cite{neupane2024momentary} fulfills both of these needs. Participants were prompted when physiological events were detected via wrist-worn wearables, using a stress detection model~\cite{CuesHub}. This ensured that stressors were captured as soon as they occurred. A confirmation by the participant ensured that only valid stressors were added to this dataset.
%, and any false detection by the model did not impact the dataset. 
Even though only a subset of detected events were used to generate prompts to limit participant burden, they were distributed to events of all kinds (low arousal to high arousal). Large amount of data, collected over 100 days from 68 participants who came from various age groups and occupations, ensured that the dataset consists of stressors reported at all times of the day and covers a wide spectrum of stressors. 
{A total of 24,459 events were responded to by participants, which consisted of 9,797 valid stressors (a clear stressor description that unambiguously identified a source of stress). }

\subsubsection{Stressor Classification using LLM}
To manage the large number and variability of unique stressors collected in the MOODS dataset (N>1500), we implemented a two-step process to map free-form stressor reports into stressor categories. First, we employed a human-in-the-loop harmonization procedure augmented by GPT-4o via prompt engineering~\cite{OpenAI2024GPT4}. This step involved consolidating semantically similar stressor phrases while preserving their underlying diversity, reducing the total number of unique stressors from over 1500 to 1120 harmonized stressors.
Second, because the field dataset lacked predefined category labels, we leveraged the Stress Annotated Dataset (SAD)\cite{mauriello2021sad}, which provides diverse examples across nine stressor categories at both phrase and sentence levels. To minimize human bias in category-level assignments and leverage the generalization capabilities of large language models, we fine-tuned a LLaMA 3 (8B) model\cite{llama3modelcard}  using LoRA~\cite{hu2022lora} on the SAD dataset. This fine-tuned model was then used to classify the harmonized stressors into nine categories.
Upon close inspection of the \textit{Other} category, we identified three more stressor categories that warranted their own categories. This refinement resulted in a final taxonomy of twelve stressor categories used in our analysis, as referenced in Table~\ref{tab:data_vs_model_all}.

\subsection{Asymptotic Model of Stressor Frequency}
Even though MOODS is a more suitable dataset of stressors, estimating the latent frequency of daily stressors still poses several challenges.  It requires correcting for biases in prompt timing, frequency, and user response behavior. A naive approach might attempt to extrapolate stressor rates for each hour of the day, but such hourly modeling would require dense sampling and careful correction for temporal biases, given that stressor likelihood is not uniform across the day. Instead, we shift the modeling paradigm from time-based extrapolation to prompt-based simulation and modeling, where we estimate the number of stressors as a function of increasing prompt response frequency. This framing enables us to investigate the saturation behavior in stressor reporting as responses increase. The next section describes the design of this simulation framework.

\subsubsection{Simulation Framework}

To enable estimation of the latent frequency of daily stressors, a simulation algorithm must meet several key objectives. First, it should remove dataset-specific prompting biases introduced by the original prompting algorithm. Second, it must generate a full-day sequence of physiological events as would be detected by an AI model. Third, each event must be assigned a stress likelihood score. Fourth, the simulation should capture the variability between people in how the AI model assigns these scores. Fifth, it must account for individual differences in how people respond to prompts based on varying levels of stress likelihoods. 
Finally, the simulation should reflect the complexity of the real world by modeling the interaction between physiological events, stress scores, and the response behavior of the participants. We now describe how our simulation algorithm meets these objectives.

\subsubsection{Addressing Prompting Bias}\label{sec:bias}

The MOODS study employed a specific prompting strategy to decide which detected events triggered a prompt. Each day, it selected one event with $\leq 25^{\text{th}}$ likelihood percentile, two events in the ($25^{\text{th}}, 75^{\text{th}}$] percentiles, three events in the ($75^{\text{th}}, 95^{\text{th}}$] percentiles, and all events with $>95{^\text{th}}$ percentile, where percentiles were computed using person-specific data. To remove this selection bias, we stratify all rated events into 20 buckets of 5-percentile likelihoods each. When generating an event for a specific time point, it is selected from any of these 20 buckets with equal probabilities, which removes the original sampling bias in the MOODS data (see Section~\ref{sec:simulation}). For the simulation, we used data from 68 participants who had rated at least one event in each of the 20 buckets.

\subsubsection{Percentile Threshold for Candidate Prompts} 
\label{sec:design_guidelines}

Selection of candidate prompts faces a significant challenge due to wide variability in stress likelihood scores for large between-person variability in physiology and physiological responses (base heart rate and how it changes during stressful moments~\cite{smets2018large}) and biases in the AI model.
It can result in widely varying performances for different participants~\cite{smets2018large}. Therefore, instead of using a population-wide uniform threshold~\cite{sarker2017markers}, we construct \emph{person-specific percentiles} to have a more uniform distribution of prompts across participants.
We define the following parameters to determine the percentile threshold for selecting candidate prompts.
\begin{enumerate}[label=(\roman*)]

   \item \emph{Expected watch-wearing hours per day ($\omega$)}

    \item \emph{Expected number of potential prompts per hour ($\eta$)}
   
     \item  \emph{Response frequency ($k$)}: It depends on the response rate expected in a study which we denote by $\alpha$. To achieve $k$ responses per day, the study will need to deliver $k/\alpha$ prompts. 

    \item  \emph{Percentile Threshold ($\rho$)}: Prompts in the top $(100*\rho)^\text{th}$ percentile (or $\rho$ percent of prompts) are sent to the participants. 

\end{enumerate}
These parameters are related via the following equation which can be used to determine any, given the others.
% \vspace{-5pt}
\begin{equation}\label{eq:random_prob}
\rho\times(\eta \times \omega) = \frac{k}{\alpha} ; 0 < \frac{k}{\alpha} \leq (\eta \times \omega);  0 < \eta, \omega 
\end{equation}
The right side in Equation~\eqref{eq:random_prob} represents the number of daily prompts, and the left side represents the number of candidate prompts expected per day that meet the threshold. The first inequality shows that the number of daily prompts ($\frac{k}{\alpha}$) can not exceed the total number of potential prompts ($\eta \times \omega$). Further, even though $0 < \rho\leq 1$ by definition, $\frac{\frac{k}{\alpha}}{\eta \times \omega} \leq \rho\leq 1$ from Equation~\eqref{eq:random_prob}. 

A study can select a value for $\rho$ within its valid range to determine the response frequency $k$, or alternatively, specify a desired $k$ to compute the corresponding $\rho$. The choice of $k$ is often guided by the target number of stressors the study aims to capture per participant per day. We now explore how varying $k$ influences the number of stressors recorded.
We assume that a study has an estimation of parameters $\eta$, and $\omega$.

\subsubsection{Data-Driven Simulation}\label{sec:simulation}

To estimate the frequency of daily stressors for different response frequencies ($k$) from 1-12, we do the following: 

\begin{enumerate}[leftmargin=*]
    \item \textbf{Estimating Daily Stressors}
    \begin{itemize}[leftmargin=*]
    {
        \item Total expected prompts per day:} $E_d = \eta \times \omega$.
        We set $\omega = 12$, $\eta = 2.5,$ following~\cite{liao2018just} and the MOODS study, respectively.
        \item For each participant, we simulate 1,000 typical days with $12\times2.5=30$ prompts per day.
        \item To generate each prompt in a day, we randomly sample one of 20 buckets (Section~\ref{sec:bias}) and an event from the bucket.        
        \item We compute the percentile threshold ($\rho$) using Equation~\ref{eq:random_prob} to compare it with the percentile score of the intensity of the sampled event to determine its eligibility for prompting.
    \end{itemize}
    
    % \item \textbf{Assessing Participant Validation}
    \begin{itemize}[leftmargin=*]
        \item For each delivered prompt \( E_p \), the presence or absence of a participant-reported stressor (as recorded in the MOODS dataset) is used to compute overall response efficiency.
    \end{itemize}
    
    % \item \textbf{Simulating Across the Population}
    \begin{itemize}[leftmargin=*]
        \item Repeat the process for all participants to estimate an average number of daily stressors (obtained by multiplying response efficiency with responses
per day).
    \end{itemize}

    \item \textbf{Normalizing for Prompt Fatigue}
\begin{itemize}[leftmargin=*]
    \item Response efficiency is expected to decrease as the response frequency ($k$) increases. We first estimate the base response efficiency for each $k$ using data from 68 participants, followed by applying a population-level linear mixed model (LMM) computed from the observed MOODS data.
    
    \item The population LMM fit $b=0.353 (\pm 0.172)$ and 
    $m=-0.007$
    $(\pm 0.01)$
    % ($b=0.353 (\pm 0.172)$ and $m=-0.007 (\pm 0.01)$) 
    is used to scale the base response efficiency up or down for each $k$, relative to $k = 3.89$ (the average response per day in the MOODS Study).
    
\end{itemize}
\end{enumerate}

We repeat the simulation process to estimate category-wise frequency of stressor reports by determining which stressor category \(C_i\) the prompted prompt \(E_p\) belongs to, if a stressor is reported. We restricted our analysis to events reported between 8:00 AM and 8:00 PM in the MOODS dataset, representing a 12-hour day.

\subsection{Modeling Diminishing Returns and Saturation Point}
The daily count of reported stressors is a function of response frequency, which is contingent on the number of prompts administered to participants. However, increasing the number of prompts and therefore responses may not lead to a proportional rise in reported stressors due to the factors outlined in~\ref{sec:problem}. Typically, such trends are modeled using one of three approaches: linear, hill, or exponential models~\cite{Sellforte,tellis2006modeling}. Given our hypothesis that marginal benefit diminishes with increasing responses, we model this relationship using an asymptotic exponential function, which captures the expected pattern of a rapid initial rise followed by gradual saturation, ultimately approaching an upper bound that reflects natural limits on reported stressors. The following equation represents the relationship between response frequency and stressors:
% \vspace{-5pt}
\begin{equation}
    y(k) = S \left(1 - e^{-ak}\right)
    \label{eq:diminishing-returns}
\end{equation}

\begin{itemize}[leftmargin=*]
    \item $k$ is the number of prompt responses, $y(k)$ is the expected number of reported stressors, and $S$ is the maximum possible stressors (saturation point) per day.
    % \item $a$ controls the {rate of growth}, modeling the decline in stressor reporting probability as responses increase.
    % \item   $a$ controls the {rate of growth}, with higher values indicating faster growth and earlier saturation.
    % \item $a$ controls how many responses are needed to capture a specific proportion of stressors of this type (e.g., half); higher values mean fewer responses needed to capture a majority of these stressors.
    \item $a$ controls the number of responses required to capture a certain proportion of stressors (e.g., 50\%). Higher values of $a$ correspond to faster saturation, meaning fewer responses are needed to capture most stressors of that type.

\end{itemize}

As \( k \to \infty \), the exponential term \( e^{-ak} \to 0 \), and the expected number of stressors per day converges to the saturation point:
\vspace{-5pt}
% \[
% \lim_{k \to \infty} y(k) = S
% \]
\begin{equation}
    \lim_{k \to \infty} y(k) = S
    \label{eq:asymptote}
\end{equation}
This asymptotic behavior reflects the principle of diminishing returns; even with an infinite number of prompt responses, the number of reported stressors will not exceed \( S \). The model thus aims to capture the nonlinear relationship between response frequency and reported stressors, ensuring a realistic estimation of the upper bound on daily stressor experiences.

\subsubsection{Weekly-Level Frequency}\label{sec:week}
For better interpretation, we also report our findings at the weekly level stressor frequency by multiplying the saturation value \( S \) obtained from the exponential model by 7, reflecting maximum stressor reports over a week. 
To compare with partial observations from the MOODS data, we also extrapolate weekly stressor counts from the dataset, for both overall and by category. In the MOODS study, participants wore a smartwatch for an average of 7.2 hours per day. We scale observed stressor counts to a 12-hour day using a factor of $12/7.2$, then multiply by 7 to estimate the number of stressors expected in a full week.

% \vspace{-10pt}

\begin{figure}[h]
    \centering
    \includegraphics[width=0.9\linewidth]{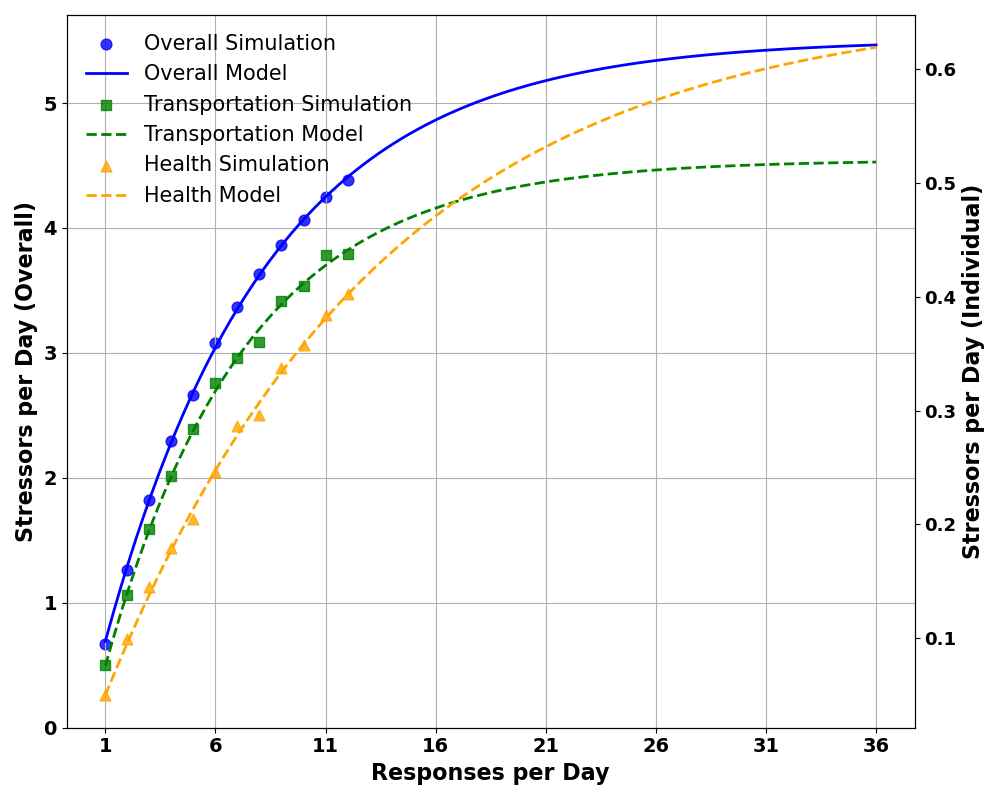}
    \caption{Simulation vs. modeled stressors per day. Scatter points represent simulated values; lines show exponential model fits. The right axis shows individual stressor category frequency; the left axis shows total stressor frequency.}
    \label{fig:exponential_fit}
    \Description{Figure shows simulation (markers)   and model (lines) results for all stressors, transportation-related, and health-related stressors for response frequency from 1 to 36 per day (1/12 to 3 responses per hour), respectively. We observe diminishing returns: increases in response frequency lead to a marginal increase in additional stressor reports.}

\end{figure}

% \vspace{-14pt}

\section{Results}

\subsection{Relationship of Response and  Stressor Frequency}

Figure~\ref{fig:exponential_fit} shows simulation (markers)   and model (lines) results for all stressors, transportation-related, and health-related stressors for response frequency from 1 to 36 per day (1/12 to 3 responses per hour), respectively. We observe diminishing returns: increases in response frequency lead to a marginal increase in additional stressor reports. This aligns with the classical \textit{Law of Diminishing Returns}~\cite{shephard1974law}, where increasing sampling frequency ($k$) leads to progressively smaller gains in observed stressors due to reporting fatigue and the finite number of real stressors experienced in a day.

For the overall population, the model saturates at approximately $S = 5.39$ stressors per day, indicating a practical upper bound on the number of stressors that can be captured, even with very frequent prompting and response. A similar pattern of diminishing returns and saturation is observed across individual stressor categories as well (Table~\ref{tab:data_vs_model_all}). While increasing response frequency initially leads to higher stressor capture, the marginal gains decrease over time, and the benefit eventually plateaus.

\subsection{Stressor Distribution Across Categories}

Table~\ref{tab:data_vs_model_all} summarizes weekly stressor frequencies estimated from extrapolated data (Data*) and model (Model) across 12 stressor categories (see Section~\ref{sec:week}). Saturation was observed across all categories, indicating sufficient data to model each category of stressors. Substantial differences in saturation levels were observed across categories. 
The top three (\emph{Work-related}, \emph{Health, Fatigue, or Pain}, and \emph{Transportation}) together account for more than 50\% of the cases, indicating how disproportionately they affect people.

Interestingly, although \textit{Transportation} stressors are more frequent than \textit{Health}-related stressors in extrapolated observed data (2.44/week vs 1.92/week), the model-estimated latent frequency for health-related stressors is higher. \textit{Transportation} stressors tend to occur briefly in the morning and evening, leading to faster saturation at lower response frequencies. In contrast, health-related stressors can appear randomly and often persist throughout the day, making them strong candidates for repeated capture as the response frequency increases. The model, as evident in Figure~\ref{fig:exponential_fit}, captures these dynamics, showing a slower decline for \emph{health-related} stressors versus a more rapid saturation for \emph{transportation}. This demonstrates the utility of model-based estimation in uncovering latent patterns in daily experiences, extending beyond sparse observations.

\vspace{-5pt}

\begin{table}[h]
\caption{
\textbf{Model} column reports the exponential model parameters: $S$ (saturation per day) and $a$ (rate of growth), along with the expected number of weekly stressors ($7 \times S$). Extrapolated weekly stressors observed in the dataset are denoted as \textbf{Data*}.
}
\label{tab:data_vs_model_all}
\centering
\begin{tabular}{|l|ccc|c|}
\hline
\textbf{Stressor} & \multicolumn{3}{c|}{\textbf{Model}} & \textbf{Data*} \\
 & \textbf{$S$} & \textbf{$a$} & {Weekly} &  \\
\hline
Work & 1.76 & 0.12 & 12.32 & 7.22 \\
Health, Fatigue, or Pain & 0.59 & 0.09 & 4.13 & 1.92 \\
Transportation & 0.55 & 0.15 & 3.85 & 2.44 \\
School & 0.42 & 0.12 & 2.94 & 1.37 \\
Emotional Turmoil & 0.40 & 0.10 & 2.80 & 1.66 \\
Social Relationships & 0.39 & 0.13 & 2.73 & 1.84 \\
Family Issues & 0.22 & 0.16 & 1.54 & 0.98 \\
Everyday Decision Making & 0.20 & 0.20 & 1.40 & 1.26 \\
Playing games/sports & 0.12 & 0.31 & 0.84 & 0.87 \\
Chores & 0.07 & 0.17 & 0.49 & 0.37 \\
Financial Problem & 0.03 & 0.11 & 0.21 & 0.10 \\
Other & 0.79 & 0.17 & 5.53 & 3.38 \\
\textbf{All Stressors} & 5.39 & 0.14 & 37.73 & 23.4 \\
\hline
\end{tabular}
\end{table}

\section{Discussion}

\subsection{Clinical Utility of Stressor Frequencies}
Latent stressor frequencies offer a normative reference point for interpreting daily stress exposure, supporting more informed clinical decision-making. For example, the model estimates an average of 12 work-related stressors per week in the general population. If a patient reports significantly more—say, 20 or more—this deviation from the norm could signal elevated psychosocial burden, prompting further clinical evaluation. 
Such population-informed baselines allow clinicians to assess whether a patient’s stressor load is within typical bounds or indicative of heightened risk. By contextualizing individual reports against expected frequencies, these estimates can guide triage decisions, inform treatment prioritization, and support tailored intervention strategies. In this way, our work enables the usage of observational data from ubiquitous sensing to clinically actionable insights.

\subsection{Design Implications for JITAIs}

Previous research in mobile health has developed methods to forecast the number of stress events to decide when to administer JITAIs under tight prompt budgets~\cite{liao2018just}. These methods emphasize balancing timely support with user burden by assigning prompt probabilities based on the number of risk events expected in a day.
Our work contributes to this foundation by modeling the frequency of specific stressors using fitted parameters $(S, a)$. This allows the intervention probabilities to be tailored to each type of stressor.
By enabling stressor-specific probability estimates, our method supports dynamic and personalized allocation of a limited intervention budget, moving beyond binary stress detection and generic prompts. 
This finer-grained approach helps to address the limitations of one-size-fits-all interventions. Prior studies show that users prefer stressor-specific support~\cite{howe2022design}, and that only a minority (20\%) of physiologically detected stress events require intervention~\cite{neupane2025wearable}. Adapting JITAI schedules to individual stressors while accounting for moments when interventions are desired can improve effectiveness, reduce fatigue, and support more sustainable, population-aware intervention strategies.

\subsection{Estimating the Economic Cost of Stress}

To estimate the economic impact of stress and burnout, two core components are needed: (1) how frequently individuals experience stress, and (2) how often they experience different types of stressors. Our approach provides these foundational inputs by modeling stressor-specific frequencies and fitted daily reporting parameters $(S, a)$, which can be used to construct realistic population-level stress timelines.
These stressor frequencies can be integrated into agent-based simulations or Markov models, such as those used in recent work on burnout cost modeling~\cite{martinez2025health}. 
Moreover, this framework enables estimating hidden costs of stress in domains often overlooked—for example, stress due to transportation can now be evaluated in terms of stress-related health and economic burden. By embedding our stressor-frequency modeling into health-economic simulations, organizations and policymakers can conduct scalable evaluations of interventions without expensive data collection.

\section{Limitations}
While this paper offers valuable insights into the relationship between response frequency and stressors, several limitations warrant consideration.
First, our analysis is based solely on the MOODS dataset, which may limit generalizability to other populations or contexts. Second, we use data collected from a single commercial stress detection model; results may differ with alternative models due to varying features and thresholds. Third, the simulation does not account for external factors such as demographics or contextual influences that affect self-reporting.
Despite these limitations, our work provides a foundational framework for estimating stress reporting limits and informing prompting strategies. Future research can validate our findings across other datasets and stress detection models to strengthen generalizability.

\section{Conclusion}

How many times people experience stress of different kinds in real life is a fundamental question with implications across physical health, mental health, the workplace, relationships, among others. By using a new dataset and novel modeling approaches to address sampling biases and sparsity, this work showed that people experience around 5.39 stressors per day. This has implications for the research methods used to study stress and develop stress interventions. Understanding the stressor frequency norm in a general population is only the start. People in different stages of life, in different occupations, in different geographies, in different sociodemographics, living with different health conditions, etc., are likely to have different norms. The stressor frequencies specific to them can now be estimated and become a new measure to understand the stressor burden differences across various groups.

\begin{acks}
Research reported here was supported by the National Institutes of Health (NIH) under award P41EB028242. It was also supported by the National Science Foundation (NSF) under awards ACI-1640813 and CNS-1822935. The opinions expressed in this article are the authors' own and do not reflect the views of the NIH or NSF.
\end{acks}

\bibliographystyle{ACM-Reference-Format}
\balance
\bibliography{bibliography}

\end{document}